\documentstyle[11pt]{article}
\newcommand{\beq}{\begin{equation}}
\newcommand{\eeq}{\end{equation}}

\newcommand{\g}{\gamma}

\newcommand{\rh}{\rho}
\newcommand{\sg}{\sigma}

\renewcommand{\l}{\lambda}

\renewcommand{\b}{\beta}
\renewcommand{\a}{\alpha}
\newcommand{\n}{\nu}
\newcommand{\m}{\mu}
\newcommand{\th}{\theta}

\newcommand{\z}{\zeta}

\newcommand{\e}{\epsilon}

\begin{document}
\topmargin 0pt
\oddsidemargin 5mm
\headheight 0pt
\headsep 0pt
\topskip 5mm

\begin{flushright}
BCUNY-HEP-93-1\\
\hfill
March 1993
\end{flushright}

\begin{center}
\hspace{10cm}

\vspace{48pt}
{\large \bf
THE SPACE-TIME MANIFOLD AS A CRITICAL SOLID }
\end{center}

\vspace{10pt}
\begin{center}
{\em Dedicated to the memory of Robert E. Marshak}
\end{center}

\vspace{30pt}

\begin{center}
{\bf Peter Orland\raisebox{.6ex}{*}}\footnote{\raisebox{.6ex}{*}
Work
supported by PSC-CUNY
Research Award Program grant nos. 662467 and 663368.}

\vspace{42pt}

The City
University of New York, \\
Baruch College, \\
17 Lexington Avenue, \\
New York, NY 10010, U.S.A.\\
orlbb@cunyvm.bitnet\\
orland@kronecker.baruch.cuny.edu\\
orland@nbivax.nbi.dk

\end{center}

\vspace{25pt}

\begin{center}
{\bf Abstract}
\end{center}
It is argued that the problems of the cosmological
constant, stability and renormalizability
of quantum gravity can be solved if the space-time manifold is
not fundamental, but arises through spontaneous symmetry breaking. A
``pre-manifold" model is presented in which many
points are connected by
random bonds. A set of $D$ real numbers is assigned
to each point. These numbers are coupled between points connected
by bonds. It is then found that the dominant configuration
of bonds is a flat $D$-dimensional manifold, on which there
is a massless matter field. Adjusting the parameters of the
model leads to fluctuations which at large
distances describe quantized
massless gravity if $D=\;4$, $6$, $8...$. These
fluctuations do not destabilize
the manifold. An approach to include
Lorentzian signature is presented.

\vfill

\newpage

\section{Introduction}

Any attempt to quantize gravity faces challenging
questions. How can the theory be renormalized, but still give Einstein's
equations in the infrared? In the Euclidean formulation, how
can the conformal
instability of the action be cured? If the instability can indeed be
cured, why
don't space-time fluctuations
lead to a creased or branched-polymer phase
\cite{jonsson} \cite{surfaces} like
that in random surface
models, destroying the interpretation of the manifold as space-time? What
makes the manifold nearly flat, in other words, why should the
cosmological constant be zero?

The most popular approach to quantum gravity has been through supergravity
theories and, more recently, superstrings \cite{green+schwarz}. Another
program, begun by Ashtekhar, is to canonically
quantize the theory in a new way \cite{ash}.  Unfortunately
difficult, perhaps insurmountable technical difficulties face such
programs.  Numerical methods have been used
to study functional integrals for lattice
Euclidean gravity
\cite{ambjorn}. It seems that a continuum limit exists, but
one in which the vacuum curvature does not
vanish \cite{amb+jur}. While anyone
working on the problem of quantum gravity needs an
optimistic temperment, no attempt has met with even qualified success.

The issue taken up here is not whether
the approaches above are mathematically
adequate to solve the problems of quantum
gravity. Rather it is that
instead of beginning by formally defining or generalizing
Einstein's theory, one should try a more
intuitive tack and look to a physical example for guidance. Such
an example exists; the melting of a solid.

In solids and monatomic incompressible liquids
it is natural to view each pair of nearest
neighbor atoms being joined a bond. A discrete
manifold is then defined by the graph of points (atoms) connected
by links (bonds). In
solids held together by covalent or hydrogen bonds, these bonds have a
genuine physical meaning (such as a shared electron). In liquids they do
not exist between different molecules; however
they are still a useful fiction. The statistical
mechanics of a liquid membrane \cite{peliti} is
described by a covariant string model \cite{polyakov} for precisely this
reason. The dominant manifolds are triangular in the partition
function. By viewing each bond as being of length $a$, a metric
is defined, thereby leading to an effective model
such as that discussed in ref. \cite{surfaces}. If the
number of points diverges at the
critical point, random surface models, and
possibly models of random simplicial lattice manifolds in higher
dimensions \cite{ambjorn}
will polymerize \cite{surfaces} \cite{polyakov2}. This
means that the manifolds collapse to
one-dimensional structures which may branch.

It is suggested here that
a gravitational partition function free of
the branched polymer disease with vanishing vacuum curvature
is not a sum over liquid
manifolds at all, but must be a sum over the manifolds
of a ``critical solid". The partition function of
an ordinary simplicial lattice crystal
solid is dominated by flat manifolds, so polymers
do not form. The variations of the
manifold are short-range in ordinary three-dimensional solids. This
means that the effective theory of gravity
is massive. It is argued here that under certain circumstances
that the solid-liquid phase boundary can
become second-order or critical. The
mass of the gravitational fluctuations in
the crystal phase near this phase
boundary can then be made as small as
desired. Furthermore, the presence
of long-range order will guarantee that
the manifold will tend to
be flat in the infrared. The sum over
manifolds is then
indistinguishable from
that of a liquid phase at distances
shorter than the correlation length. This
length is the Compton wavelength of the graviton.

In conventional models of random manifolds, it appears that crystal
formation is ruled out. It therefore seems the
most sensible way a space-time manifold
can stay approximately flat is the same way the manifold
of a solid does. The manifold
should not be put into the theory
at all in the beginning, but should
appear dynamically.

In this paper a simple model with no {\em a priori}
space-time manifold is presented. This model
consists of a set of points randomly connected by bonds. To each
point is assigned a set of $D$ real numbers. It
is found that a flat $D$-dimensional simplicial manifold can
arise spontaneously, much in the way that a crystal does. Furthermore
if $D=\;4,\;6,\;8\;...$ the melting transition can be second
order, that is the solid has a
critical region. Fluctuations
in the metric of the manifold lead to a discretized formulation
of Riemannian
quantum gravity \cite{ambjorn} coupled to a scalar field at
large distances, if the solid is
close to the melting transition. Since
the system is critical, it can be renormalized, though not by
perturbation theory. At the critical point, dimensionful
quantities scale in such a way that the cut-off can be removed, leaving
a renormalized field theory (for an
example of the renormalization of a perturbatively
non-renormalizable theory see \cite{gawed}). It was first suggested
by Weinberg that a formally non-renormalizable theory of gravity
might have a critical surface \cite{weinberg}. It is quite striking that
the smallest value
of $D$ for which a critical solid exists is $4$ (as the
criterion of renormalizable field theory implies $D\le 4$, the
space-time dimension is completely fixed). Since the fluctuations
are around a flat manifold, the cosmological constant is
automatically zero.

Fluctuations in the basic structure of space-time beyond the
Planck scale have been speculated about for some time
\cite{wheeler}. Fluctuations specifically in topology were considered
by Hawking \cite{hawking} in the context of Regge calculus
\cite{regge}. It has been suggested that Regge calculus
or a related formulation might be the physically correct
description of the space-time manifold
above the Planck mass \cite{nielsen}. Topological
fluctuations were suggested to have important dynamical
effects by Coleman \cite{coleman}. If the picture
presented here is correct, notions like topology, dimension, etc. begin
to make sense only below
the Planck scale. It is my own suspicion that critical solid formation
could arise in other pre-manifold models which are not
based on the notion of a discrete set of points, such as string
theory (in which the appearance of space-time is not yet
understood) \cite{green+schwarz} or
some other (true?) theory of gravity yet hidden from the
imagination.

A second model is discussed in which gravitation with Lorentzian
signature might arise. The motivation lies in the fact that a Lorentz
metric can be constructed from a Riemannian metric and a vector
field \cite{geroch}. Unfortunately, the validity of
the simple picture of the
spontaneous appearance of a flat manifold with gravitational
fluctuations is not as obvious as for the first model.

Some suggestions for further numerical and analytic work are
presented in the conclusion.

\section{Discretized Gravity}

The purpose of this section is to briefly familiarize the reader
with the regularized formalism of gravity discussed in references
\cite{ambjorn}, \cite{amb+jur}, \cite{nielsen}. The power of this
formalism is that there is no need for explicit coordinates. The
law of gravitation can be expressed in terms
of a network of points, guaranteeing general covariance.

Suppose $\cal M$ is
a $D$-dimensional
simplicial manifold (usually called a ``simplicial complex"
by mathematicians) \cite{s+t} of fixed
topology. This means that $\cal M$ is a network
of points (which will be called ``sites")
connected by line segments (which will be called ``links") which can
be constructed by gluing together $D$-dimensional simplices
${\bf K}_{D}$ along
their $D-1$-dimensional sub-simplices ${\bf K}_{D-1}$. If
$D=2$, triangles, ${\bf K}_{2}$,
are glued together by identifying links, ${\bf K}_{1}$, if
$D=3$ tetrahedra, ${\bf K}_{3}$, are connected
by identifying triangles, etc. If $\cal M$ has no
boundary, each $D-1$-dimensional
subsimplex is shared by two $D$-dimensional simplices. At boundaries
$D-1$-dimensional simplices are not shared.

In the original formalism of Regge \cite{regge} the network
of sites and links (Regge called the links ``bones")
is fixed and each link is assigned a variable
length. A
simplex is regular if and only if the lengths
of all its links are the same. In the alternative approach
of references
\cite{ambjorn}, \cite{amb+jur}, \cite{nielsen} the links are
instead have a fixed length $a$. The quantum functional
integral is defined as a sum over different manifolds (possibly
of
fixed topology)
$\cal M$,
\beq
Z=\sum_{\cal M} e^{-S({\cal M})}\;.   \label{1.1}
\eeq
The sum in (\ref{1.1}) becomes a sum over metrics in the continuum
limit $a\rightarrow 0$. The metric is defined to be flat inside
each $D$-dimensional
simplex, consistent with links having length $a$. Curvature can exist
on the boundaries between simplices. Each simplex has the fixed volume
it would possess in flat space.

A general coordinate invariant
action, $S({\cal M})$, should be local, at least as
$a\rightarrow 0$. It will contain covariant terms depending
on the metric and the Riemann tensor, if
coordinates are chosen. The
discrete version of the cosmological term is
\beq
S_{cosm}({\cal M}) = \Lambda \sum_{{\bf K}_{D}} V({\bf K}_{D})
\rightarrow \int\,d^{D}\,x g^{1/2}\;,  \label{1.2}
\eeq
where $V({\bf K}_{D})$ is the volume of a $D$-dimensional
simplex (a fixed constant). The discrete Einstein action
is
\beq
S_{Einstein}({\cal M})=-\frac{1}{16 m_{p}^{2}} \sum_{{\bf K}_{D-2}}
\delta({{\bf K}_{D-2}}) V({\bf K}_{D-2})
\rightarrow -\int\,d^{D}\,x g^{1/2} \,R\;. \label{1.3}
\eeq
In equation (\ref{1.3}), $\delta({{\bf K}_{D-2}})$ is the deficit angle
associated with the $D-2$-dimensional simplex ${\bf K}_{D-2}$. Terms in
the action which are quadratic in the Riemann tensor can also be
accomodated, for which the reader is referred to reference \cite{amb+jur}.

\section{A Pre-Manifold Model}

Consider
a set of points $j=1,...,N$. To each of these points a value of the
``pre-phonon field" $\phi_{j} \epsilon {\cal R}^{D}$ is assigned. Later
on, when it is explained how the space-time manifold forms, it will
become clear that $D$ is in fact the dimension of this manifold. To
each pair of points $j \neq k$ is assigned
a ``link variable" $s_{j\,k}=s_{k\,j}=0,1$. The partition
function of the model
is
\begin{eqnarray}
Z&=     &\sum_{N=0}^{\infty} \z ^{N} [\prod_{j=1}^{N} \int\;d^{D}\phi_{j}]
         [\prod_{j<k} \sum_{s_{j\,k}=0,1} g(s_{j\,k})]    \nonumber \\
 &\times&\exp-[\frac{c}{2}\sum_{j<k}s_{j\,k}(\phi_{j}-\phi_{k})^{2}
+K\sum_{j<k} \th (R-|\phi_{j}-\phi_{k}|)]\;\;, \label{2.1}
\end{eqnarray}
where $\th$ is the usual step function, $g(0)=1$
and $g(1)=g$, $\z$, $c$, $K$ and $R$ are all real
constants. The link variable $s_{j\,k}$ is quite similiar to that
used in a non-perturbative formulation of strings \cite{giles}. The number
of points, $N$, is eventually taken to infinity.

In (\ref{2.1}) a natural metric exists. The length if each link is defined
to be $a$. The the distance
between any two points is defined to be equal to the length of the
shortest path connecting them; if no such path exists, the distance
between them is infinite. Thus, even if the set of points together
with this distance function
do not form a manifold, they satisfy the conditions of a metric space. It
for this reason (\ref{2.1}) is called a ``pre-manifold" rather
than a ``pre-geometric" model.

Why should this be a theory of gravity? Specifically: \begin{enumerate}
\item Is there a saddle point of the sum in (\ref{2.1}) for which
$s_{j\,k}$ describes a discrete manifold of small constant curvature, i.e.
a nearly regular simplicial lattice? If so, there will be a massless covariant
phonon field
$\psi_{j}=\phi_{j}-<\phi_{j}>$ on this discrete space-time.
\item Are the fluctuations in $s_{j\,k}$ near this saddle point correctly
describing quantum gravity in the infrared? Allowed fluctuations can be
dislocations and
disclinations, but there should be no tearing of the lattice. Since the
graviton is massless, these fluctuations must have an infinite correlation
length. There must be no instabilities destroying the manifold.
\end{enumerate}

It is not hard to see, using a little intuition garnered from the physics
of crystal formation that the answer to the first question is
{\em yes} under a wide
variety of choices of the constants, provided $D>2$. Furthermore, $D$
turns out to be the dimension of the dominant manifold. I will also
argue on the basis of this theory that the answer to the second question
is also affirmative, provided
the couplings are tuned correctly and $D$ is even. The resulting
theory of gravity can therefore be only in $D=4,6,8,...$ dimensions. The
cosmological constant is zero, as the saddle point is a flat manifold. The
fluctuations are not strong enough to destroy the manifold, so there is
no conformal instability, nor is there a pathological branched
polymer phase. Furthermore
the theory exists at a critical point, and so is well behaved in
the ultraviolet.

\section{Flat Manifold Formation}

To get some understanding of what is going on, it is easy to sum over
$s_{j\,k}$ in (\ref{2.1}) to get
\beq
Z=\sum_{N=1}^{\infty} \z^{N} [\prod_{j=1}^{N} \int\;d^{D}\phi_{j}]
\exp-[V(|\phi_{j}-\phi_{k}|)]
\;\;, \label{2.2}
\eeq
where
\beq
V(|\phi_{j}-\phi_{k}|)=-\log[1+e^{-\frac{c}{2}(\phi_{j}-\phi_{k})^{2}}]
+K \th (R-|\phi_{j}-\phi_{k}|) \;\;. \label{2.3}
\eeq
Now $V(|\phi_{j}-\phi_{k}|)$ can be interpreted as
a two-particle potential for a system of
atoms. Assume that the repulsive core radius, $R$, is
smaller than $c^{-1/2}$. This potential is attractive
for  $R<|\phi_{j}-\phi_{j}|<c^{-1/2}$, but quickly falls off to zero for
$|\phi_{j}-\phi_{j}|>>c^{-1/2}$. Thus
$Z$ can be interpreted as the partition
function of such a collection of atoms with $\b=1/KT=1$.

If the parameters $g$, $c$, $k$, $R$ are adjusted properly a crystal
phase should
form in which the points
are close together in $\phi$-space. For other choices of these parameters
thermal fluctuations will destabilize such a crystal
forming a liquid. For the application
to quantum gravity, it is desirable that when the crystal forms, it
defines a regular
simplicial lattice (this defines a manifold with a flat
metric according to the discussion
of section 2). In one, two, three, four or
five dimensions, the most ``close packed"
arrangements of spheres, called
$L_{1}$, $L_{2}$, $L_{3}$, $L_{4}$ and $L_{5}$, respectively, are
all simplicial
lattices \cite{close-pack}. More precisely, if a site is placed at the center
of each sphere, and whenever two spheres touch a
link connects their respective
sites, the resulting lattice of sites and links is periodic and
built entirely
of $D$-simplices. There is an elementary prescription
for constructing $L_{3}$, $L_{4}$ and $L_{5}$. One simply takes a
regular square lattice of points $(n_{1}, n_{2},...,n_{D})$, where
$n_{\mu}$ is an integer, and
removes all the points
for which $n_{1}+...+n_{D}$ is an odd integer. Then one draws a sphere of
radius $1/\sqrt 2$ around each of the remaining points. The reader
can easily verify that the maximal number of spheres each of which
touches the others is $D+1$. However, in
higher dimensions, there are other lattices
which are more close packed. For $D\le 5$
it must be true that for $R$ less
than $c$, the simplicial lattice is
a solution of lower free energy than the less closely packed lattices. It
is probably also the case that simplicial lattices could form
for $D\ge 6$, but verifying this would take some work. Notice
that this scenario can work only if $D>2$, since long-ranged
crystalline order is forbidden for $D \le 2$ by Peierls' theorem
\cite{peierls}.

To actually determine the range of parameters for which a solid
crystal forms with a simplicial lattice will require an application
of density functional theory \cite{rama}. I
intend to apply this technique in a
later paper.

Now if a simplicial
crystal configuration
arises spontaneously in (\ref{2.1}), then the dominant configuration
of $s_{j\,k}$ will describe a D-dimensional simplicial
manifold. The point is that $s_{j\,k}$ is strongly correlated
with the separation $|\phi_{j}-\phi_{k}|$, i.e. for $j<k$ and
$l<m$, the correlation between the two behaves the following way:
\beq
C_{j,k;\; l,m} \equiv \; <(\phi_{j}-\phi_{k})^{2} s_{l\,m}> \;
\approx \; \Phi^{2}
\delta_{j,l}\delta_{k,m} \;\;.    \label{2.5}
\eeq
This means if $j, k$ are not nearest neighbors in $\phi$-space, hence
separated by $|\phi_{j}-\phi_{k}|\approx f$, $f>>c^{-1/2}$, it
is much less probable that $s_{j\,k}=1$ (by
a factor of $e^{-cf^{2}}$) than
$s_{j\,k}=0$, while if $j,k$ are nearest neighbors, it is more
probable that $s_{j\,k}=1$ than $s_{j\,k}=0$. This means that nearest
neighbors in $\phi$-space
are nearly always joined by links, whereas other pairs of
sites are almost never joined by links.

The massless Goldstone Boson of the sponaneous
symmetry breaking in $\phi$-space
is $\psi_{k}=a^{d/2-1}(\phi_{k}-<\phi_{k}>)$, where $a$ is the length of
a link. This can be thought of as a phonon
field in the crystal.

To summarize, it has been shown that the model (\ref{2.1}) is
dominated by a flat simplicial manifold under certain choices
of the couplings, provided the number of components of $\phi$, namely
$D$,
is greater than two. This
manifold has formed through spontaneous breaking of translation
and rotation symmetry in $\phi$-space. The resulting manifold has a dimension
equal to $D$.

\section{Metric Fluctuations and the Critical Solid}

Showing that a flat simplicial manifold
forms is a hollow victory. It is
neccesary to see what the nature of the
fluctuations around this manifold
are. What is desirable is that these
fluctuations do not shear or form cracks
in the crystal, so that the effective theory resembles one
the formulations of gravity discussed by
Lehto et.al.\cite{nielsen}, Ambj{\o}rn et. al. and Migdal
et. al. \cite{ambjorn}.

Lattice fluctuations around the flat simplicial manifold are
made by removing or adding links. They are of
two types, which will be referred to as ``geometric" and ``non-geometric"
fluctuations. The latter
include
cracks or holes (naked singularities?), local
lattice connectivities not consistent with the lattice being a manifold
(see for example the book by
Seifert and Triefal \cite{s+t}) as well as $D$-dimensional
polyhedra which are not
simplices. The former are fluctuations which change deficit angles
but preserve the simplicial charater of the lattice.

In order for cracks or holes not to form in a lattice, it must be able
to flow like a liquid at short distances. Yet
at the same time, translation invariance must
be broken and the lattice must be rigid macroscopically. The only way for
this to be the case is if the system is near a second-order phase boundary
between the solid and liquid phase. In other words the crystal is a critical
solid and the correlation length approaches infinity. Thus if a theory of
gravity arises from the model (\ref{2.1}), it will automatically be a
a massless theory of gravity. This rules out the possibility of asymptotically
free theories, which are massive in the infrared. The vacuum expectation
value of the Riemann tensor must be zero. Therefore the
effective cosmological constant must vanish. The only possible form for
the action describing the fluctuations of the simplicial manifold in
the continuum limit is
\beq
S=\frac{1}{16m_{p}^{2}}\int\,d^{4}x\;\sqrt{g}( -R
+M^{\a \b \g \e}_{\m \n \l \sg}
\;R_{\a \b \g \e}
\,R^{\m \n \l \sg}+O(R^{4}))+\int \,d^{4}x\; \sqrt{g} \partial_{\m} \psi
\partial^{\m} \psi   \;\;,   \label{2.6}
\eeq
where $m_{p}$ is the Planck mass, and the
constant tensor $M$ is numerically small compared to $m_{p}^{2}$, and
far from the perturbative fixed points of the $R^{2}$ theory (at
which the theory is asymptotically free \cite{stelle}). Notice
that the second term stabilizes the conformal instability of the first
term. Beyond the second-order phase transition, $M=0$ and the manifold is
no longer stable, becoming a liquid.

Under what circumstances can the continuum action be of the form
(\ref{2.6}), i.e.
the solid-liquid transition be second order? It
is well known that this transition is always first order
in three dimensions. The reason for this is the following. The natural
order parameter is the density
\beq
\rho({\phi})=\sum_{j} \delta^{D}(\phi-\phi_{j})\;,
\eeq
which can be approximated by a smooth function. The
Ginzburg-Landau free
energy is of the form \cite{anderson}
\begin{eqnarray}
F&=&\int\frac{d^{D}k}{(2\pi)^{D}} \a(k) \rh(-k) \rh(k) \nonumber \\
&+& \int\frac{d^{D}k_{1}}{(2\pi)^{D}}...\int\frac{d^{D}k_{D}}{(2\pi)^{D}}
\;t(k_{1},...,k_{D}) \delta^{D}(k_{1}+...+k_{D})
\;\rh(k_{1})...\rh(k_{D})\;+... \;\;, \label{2.7}
\end{eqnarray}
where
\beq
\rh(\phi)=\int\frac{d^{D}k}{(2\pi)^{D}}\rh(k) e^{ik\cdot \phi} \label{2.8}
\eeq
The reason the next to leading order term is of order $\rh^{D}$ is that the
crystal must defined by $D$ different wave vectors which
add up to zero. Thus the transition is second order if $D$ is even and
first order if $D$ is odd. Therefore, in order for a continuum theory
of gravity to arise for small $a$, the integer $D$ can only be an even
integer greater than two, $D=4,\; 6,\; 8,...$. It is
rather nice that the smallest number of
possible dimensions is four.

\section{Boundary Conditions}

Thus far, boundary
conditions in $\phi$-space and those of the
manifold in physical space have not been discussed. If the range
of $\phi$ is restricted to a finite box with ``hard-walls", the resulting
manifold $\cal M$ will have
a boundary with a Neumann boundary condition. This Neumann
condition results from minimizing the potential (\ref{2.3}) for
points on the boundary (just as Neumann boundary conditions appear
automatically in the continuum limit of open lattice strings
\cite{giles}) and are
\beq
n^{\mu} \partial_{\mu} g^{\a \b}|_{\partial{\cal M}}=
n^{\mu} \Gamma^{\a}_{\mu\;\gamma}
g^{\gamma \b}|_{\partial{\cal M}}\;\;. \label{3.1}
\eeq
Such boundary conditions seem rather unphysical from the point
of view of general relativity. On the other hand, by compactifying
$\phi$ on a region with no boundary will result in a compact boundaryless
space-time $\cal M$. In general, the topology of space-time above
the Planck length (though
not the metric!) will be the same as that of $\phi$-space. This suggests
that the topology of the universe at distances above
the Planck length is not an arbitrary consequence
of initial conditions, but
is determined by physical law.

\section{A Lorentzian Model}

No model of quantum gravity can be taken entirely
seriously unless the dominant
manifolds in the configuration space are of Lorentzian
signature $(-,+,+,+)$ and the sum over this space is a Feynman (rather
than Weiner) path integral. It is possible to define a Lorentzian analogue
of (\ref{2.1}), though understanding its behavior is harder, as there
is not yet a physical picture of crystallization.

A model similiar to (\ref{2.1}) will now be discussed for which spontaneous
formation of a manifold with Lorentzian signature could occur. The basic
modification is to introduce two types of links, which seperate points
by a time-like or space-like displacement. It is well known that any
manifold with a Riemannian metric $g^{+}_{\m\, \m}$
and a non-vanishing
vector field $\xi^{\a}$ can be interpreted
as a manifold with a Lorentz metric given by \cite{geroch}
\beq
g_{\m \,\n}=g^{+}_{\m\,\n}-
2\frac{g^{+}_{\m\,\a}\xi^{\a}g^{+}_{\n\,\b}\xi^{\b}}{g^{+}_{\rho\,\sg}
\xi^{\rho}\xi^{\sg}}     \label{4.1}
\eeq
Then the vector field $\xi^{\a}$ is time-like. The converse, that
any manifold which admits a Lorentz metric also admits
a time-like vector field, is obvious.

The model now has a three-valued link variable $s_{j\,k}=-1,0,1$ between
points $j$ and $k$. As before, $s_{j\,k}=0$ means that no link connects
these two points. If $s_{j\,k}=-1$, the link is time-like, while if
$s_{j\,k}=1$ the link is space-like. The difference between a
time-like and a space-like link is in the sign of the coupling between
$\phi_{j}$ and $\phi_{k}$. A time-like vector field is defined on the
collection of points by requiring that exactly two time-like link
are connected to each point, i.e., for any point $j$
\beq
\sum_{k=1}^{N} s_{j\,k} (s_{j\,k}-1)=4\;\;. \label{4.2}
\eeq
The model is then
\begin{eqnarray}
Z&=    &\sum_{N=0}^{\infty} \z^{N} [\prod_{j=1}^{N} \int\;d^{D}\phi_{j}]
        [\prod_{j<k}\; \sum_{s_{j\,k}=-1,0,1} g(s_{j\,k})]
        [\prod_{j}\delta(\sum_{k} s_{j\,k} (s_{j\,k}-1),4)] \nonumber \\
&\times&    \exp[-\frac{c}{2}\sum_{j<k}(is_{j\,k}-\epsilon)
(\phi_{j}-\phi_{k})^{2}
+i\sum_{j<k} \th (R-|\phi_{j}-\phi_{k}|)]
\;\;. \label{4.3}
\end{eqnarray}
Here $\delta(l,m)$ is a convenient way
of writing the Kronecker delta, $g(0)=1$
and $g(1)$, $g(-1)$, $\z$, $c$, $K$
and $R$ are all real constants. Notice that
an $i\epsilon$ prescription is built into (\ref{4.3}) to assure
convergence to a unique answer. Unfortunately, this
Lorentzian model is not some
simple analytic continuation of (\ref{2.1}). The
link variable cannot be simply summed out as before. The model is
therefore more difficult to understand physically.

\section{Conclusions}

A simple physical picture of how the space-time manifold of the
universe can arise spontaneously has been presented, and it
has been argued that this picture solves the
fundamental problems of quantized general
relativity. The discussion in this paper has been
qualitative, though the arguments are built on a
strong physical foundation.

It is important to investigate the specific form of crystal
structures which can arise for different
choices of the fundamental parameters. The best
way of studying this question is density functional theory \cite{rama}. It
should be possible to roughly locate the critical region this
way. It might also be possible to apply density functional
methods to the more interesting model ({\ref{4.3}). If so, what
would be particularly important is the global structure of the resulting
space-time, e.g. whether closed time-like curves and naked
singularities are absent
\cite{geroch}. For
the simpler model (\ref{2.2}) numerical calculations using
molecular dynamics or Monte-Carlo methods seem quite
feasible.

\section*{Acknowledgements}

I thank Jan Ambj{\o}rn and Michio
Kaku for comments on the manuscript.

\vfill












\end{document}